\documentclass[preprints,article,accept,moreauthors,pdftex]{Definitions/mdpi}

\firstpage{1} 
\pubvolume{xx}
\issuenum{1}
\articlenumber{5}
\pubyear{2019}
\copyrightyear{2019}
\history{Received: date; Accepted: date; Published: date}

\Title{Interactional and Informational Attention on Twitter}

\newcommand{\tb}[1]{\textcolor{blue}{#1}}

\renewcommand{\tb}[1]{#1}

\newcommand{\x}{\item}

\newcommand{\RT}{\mathrm{RT\,}}
\Author{Agathe Baltzer $^{1,2,\ddagger}$, M\'arton Karsai $^{1,\ddagger}$*\orcidB{} and Camille Roth $^{3,\ddagger}$*\orcidA}

\AuthorNames{Agathe Baltzer, M\'arton Karsai and Camille Roth}

\address{%
$^{1}$ \quad Univ Lyon, Inria, CNRS, ENS de Lyon, Universit\'e Claude Bernard Lyon 1, LIP UMR 5668, F-69007 Lyon, France\\
$^{2}$ \quad Mines ParisTech, France\\
$^{3}$ \quad CNRS and Computational Social Science Team, Centre Marc Bloch, UMIFRE CNRS-MAEE 14, Friedrichstrasse 191, D-10117 Berlin, Germany}

\corres{Correspondence: marton.karsai@ens-lyon.fr (M.K.); roth@cmb.hu-berlin.de (C.R.)}

\secondnote{These authors contributed equally to this work.}

\abstract{
Twitter may be considered as a decentralized social information processing platform whose users constantly receive their followees' information feeds, which they may in turn dispatch to their followers. This decentralization is not devoid of hierarchy and heterogeneity, both in terms of activity and attention. In particular, we appraise the distribution of attention at the collective and individual level, which exhibits the existence of attentional constraints and focus effects. We observe that most users usually concentrate their attention on a limited core of peers and topics, and discuss the relationship between interactional and informational attention processes --- all of which, we suggest, may be useful to refine influence models by enabling the consideration of differential attention likelihood depending on users, their activity levels and peers' positions.
}


\begin{document}


\section{Introduction}

Attention and activity seem to obey possibly conflicting dynamics. On the one hand, a wealth of results have accumulated to suggest that social attention is generally bounded, whereby humans are not able to devote their active time to more than a certain number of peers and topics \citep{hill2003social,roberts2009exploring,burke2011social,miri-limi,weng-comp}, pointing at attention as a zero-sum game constrained by limited temporal resources. On the other hand, several studies hint at reinforcement mechanisms between various attentional and activity channels, especially when comparing online with offline sociability (or social capital), which appear to be correlated \citep{well-comp,wellman-internet-increase-2001,orbach-2015-sensing,nguyen-2016-the-impact}: in this regard, attention and/or activity may breed more attention and/or activity. 

This paper aims at shedding further light on human attentional patterns by encompassing both interactional and informational aspects, and more precisely by describing the possible existence of conflicting \hbox{vs.} reinforcing cognitive constraints both in social and topical terms. To this end, we focus on a popular, observable online platform, Twitter, which features both social network and publication capabilities, thus making it possible to discuss attention from the joint perspective of interaction and information processing. Twitter allows users to constitute their own personal set of sources of which they want to follow the publications. They can also potentially dispatch these publications, thereby indicating that they indeed paid specific attention to specific users. This will enable us to make a key distinction between potential and actual attentional patterns. By doing so, we more broadly aim at describing how diverse users are in distributing their attentional resources to peers and topics: within certain cognitive limitations, such a platform also exhibits heterogeneous distributions of roles and attentional patterns. 

The next section will be devoted to a brief review of the relevant state of the art on this matter.  Section~\ref{sec:protocol} will describe the empirical data and the main definitions we use. Section~\ref{sec:social} focuses on social attention, while section~\ref{sec:semantic} introduces the notion of semantic attention and discusses its correlation with its social counterpart, which contributes to more generally address the above debates and evoke further research (section~\ref{sec:discussion}).


\section{Related work}

\noindent {\em Size of social interaction networks.}
A sizable literature shows that the number of connections in human ego-centered social interaction networks spans over several orders of magnitude, be it for scientific collaborations \citep{newm:str1}, e-mail interlocutors \citep{ebel2002scale}, content sharing platforms \citep{kumar2010structure}, online social networks \citep{mislove2007measurement}, among others. 
These studies have principally focused on the number of individuals that one may have known or have interacted with, showing that there is wide variation in human interaction \emph{potential}. When focusing on \emph{actual} interactions and, more precisely, on actively sustained interactions, 
a diverse array of studies nonetheless converges on the conception that the active core of an individual's interaction network is of bounded size. This follows from Hill \& Dunbar's seminal study on the exchange of Christmas cards \citep{hill2003social}, which shows that
individuals may actively devote their actual attention to only a portion of their potential acquaintances. Upper bounds on active connections are generally thought to be in the vicinity of a hundred people, even if this number greatly depends on various features such as, obviously, network type, connection strength thresholds and socio-demographic features, including age and gender. On the lower bound it may go to a dozen or even a handful of contacts when focusing on the most inner layers e.g., where financial support may be sought \citep{roberts2009exploring}. In terms of non face-to-face communication, human interaction capacity is shown to be similarly bounded, be it when examining phone call records \citep{miri-limi} or online social network friends \citep{burke2011social}. In other words, a variety of constraints, be they physical or cognitive, may contribute to the more or less acute reduction that an individual's social network undergoes when going from potential to actually sustained connections.

\bigskip\noindent{\em Attention dynamics on Twitter.} The issue of interactional constraints has been addressed over the recent years on various online platforms, where Twitter has increasingly served as a prototype of observable online networking processes. This scholarship has taken place within a broader questioning on attention dynamics and its possible bounds. On Twitter, social attention is subjected to a threshold similar to what is found in other contexts: for example, reciprocated (conversational) links tend to plateau after a threshold of a couple of hundreds of connections \citep{gonc-mode}. 
Semantic attention exhibits regularities at the collective level that may be interpreted as the result of underlying synchronization forces, for instance in terms of typical aggregate sigmoidal patterns of growth and decrease of the global occurrence of some topic \citep{lehm-dyna} and its burstiness \citep{sanli2015local}. 
However, a broader picture of the constraints that apply \emph{individually} to attention in this context is only partially known. The attention devoted by Facebook users to their friends tends to follow a power-law distribution, \hbox{i.e.} there is a geometrically decreasing interest as one goes down the list of a user's neighbors \cite{backstrom2011center}, while a core of about a dozen of top users appears to consistently gather a significant portion of that attention. Similar features may be found in cell phone communication patterns \citep{saramaki2014persistence}. In a more open digital public space such as Twitter, the number of unique users one pays attention to varies significantly across the platform as well as during specific events \citep{lin2014rising}. Yet, little is nonetheless known on the way Twitter users distribute and, possibly, bound their weighted attention among the sources they follow. On the semantic side too, the existence of individual limitations has not really been addressed per se, even if there are detailed accounts of the diversity of topic use in online platforms at large \citep{golder-2006-usage,wu-2007-nov} and specifically on Twitter \citep{weng-comp}. The potential combination of constraints on both the social and semantic sides remains also generally unexplored. 

\bigskip\noindent{\em Influence studies.} The modeling of social contagion is often configured as a process where the number of neighbors of a given individual matters in a uniform way, whereby all alters have an identical potential impact on ego --- be it in the canonical threshold or cascade models \citep{weng-vira} or in the so-called ``complex contagion'' models \citep{rome-diff}. In other words, whereas temporal patterns (e.g., repetition, burstiness) and topology (e.g., centrality, clustering) do generally matter, attention is rather considered equal across peers.
\tb{Attentional constraints have only recently been studied from a diffusion perspective, by examining how limits on individual processing capacities may affect the rate of propagation of information from a user to the other. For instance, \citep{hodas2012visibility} shows that there is a decreasing probability of retweeting for users who follow a larger number of people and attempt at estimating response times from observed retweets, while \citep{rodriguez2014quantifying} introduces a queuing process aimed at reflecting the sequential processing of information carried out by users in an environment when confronted with an accumulation of information in-flows. While some recent social contagion models specifically introduced the possibility of heterogeneous inter-personal influence \cite{unicomb2018threshold}, corresponding empirical characterizations appear to be still missing.}
On the whole, existing empirical studies shed an important light on the ego-centered forwarding of information at a short time-scale, yet they do not take into account a finer understanding of how users may heterogeneously distribute their attention across their neighbors or across topics.  \tb{We intend here to contribute to this question as well.}


\section{Definitions and empirical protocol}\label{sec:protocol}

\subsection{Dataset}\label{sec:dataset}
We analyze in this study a large corpus stemming from Twitter\footnote{{Twitter is a popular online news and social networking platform, which enables various types of users ---a celebrity, a news channel, you--- to create an account and briefly describe themselves, to publish messages, or ``tweets'', with restricted length and which may feature additional content (such as URLs, videos, photos), to subscribe to content generated by other users and to interact with them in various manners (republishing their posts, mentioning them, initiating conversations).}}. Our dataset has been collected over two months, between $t=\mathrm{June~6th,~2016}$ and $t'=\mathrm{August~7th,~2016}$ via the Twitter PowerTrack API provided by DataSift with an access rate of about 15\%. We focused on users who have self-declared in their account settings that they are located in the time zones GMT and GMT+1\footnote{{Note that this is not equivalent to the geolocalization of single tweets, neither to the self-reported default location of users. While such precise locational information are optional to share, time zones are requested to set for each user to match the time of their posting activity.}}. We additionally focused on tweets written in French. The dataset contains $8,233,354$ tweets and $8,698,610$ retweets. 

\begin{figure}[!t]
\centering
\includegraphics[width=.70\linewidth]{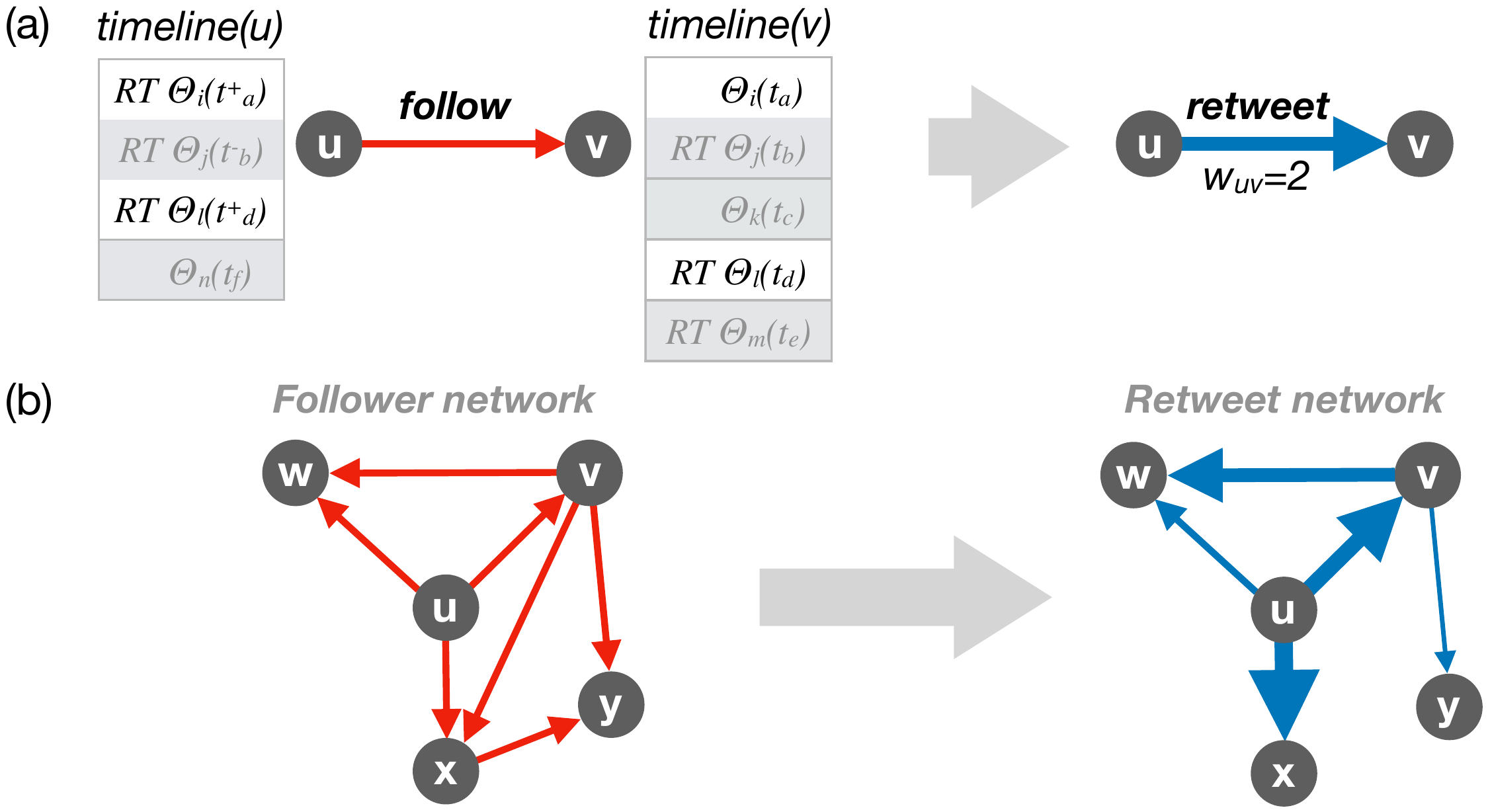}
\caption{\label{fig:schema} (a) Construction of the retweet network $R_{[t,t']}$. In the timelines of $u$ and $v$, we distinguish tweets $\Theta$ from retweets $\RT \Theta$. Given that $u$ follows $v$, we consider that $u$ retweeted $v$ if $u$ retweeted a tweet or a retweet of $v$ at a posterior time point. In this example, this includes two retweets: $\Theta_i$, which has been posted by $v$ at $t_a$ and retweeted by $u$ at ${t^+}_a>t_a$, as well as $\Theta_l$, which has been retweeted by $v$ at $t_d$ as well as $u$ at ${t^+}_d>t_d$. However, this does not include $\Theta_j$, which $u$ retweeted at ${t^-}_b<t_b$. Other tweets and retweets not common to both timelines are naturally ignored. The weight of the retweet link from $u$ to $v$ is thus $w_{uv}=2$.
(b) The retweet network $R_{[t,t']}$ is therefore a weighted sub-network of $F_t$: in both networks, the link direction from $u$ to $v$ is aligned with observed attention from $u$ to $v$. 
}
\end{figure}

\subsection{Definitions and notations}
\label{sec:def}

\bigskip\noindent{\tb{\em Focusing on retweets as attentional markers.}}
We principally focus on content flows exhibited by publication and republication dynamics, \hbox{i.e.} attention and influence dynamics between users. 
\tb{This is admittedly a proxy of a broader notion of attention. Indeed, studying attention naturally requires to define a perimeter of observation: studies using Christmas cards overlook face-to-face, e-mail or phone interactions, and vice versa. Similarly, even when focusing on just Twitter and its quite simple interaction grammar, many distinct attention channels may be considered, and thus many signals of how users pay attention. Users may read posts, their linked content (articles, videos, etc.), they may converse with other users, read posts published by users they do not follow or read the notifications they receive when they are being mentioned in other tweets, and so on and so forth. Without a comprehensive monitoring protocol able to track Twitter users and, especially, both the content they are exposed to and their actual behavior (while factoring in their varying levels of involvement on the platform: from casual to very active users) we are bound to use secondary signals.  From an attentional perspective, we thus decided to focus on user posts as \emph{visible} traces of user-centric activity. We further focus on retweets, for two main reasons:
\begin{itemize}
    \item We are interested in the cognitive filtering process that occurs between followed sources (and followees' publications) and the actual attention devoted to them. We contend that this constitutes a consistent system that enables us to properly compare what users are exposed to with what they retain. Retweeting is admittedly an ambiguous activity: it has long been considered to be influenced by a variety of temporal and individual factors, either observed \citep{suh2010want,yang2010understanding} or hypothesized \citep{zaman2010predicting}, and has been shown to range from simple acknowledgement to tentative conversation engagement \citep{boyd2010tweet}. Yet, it also positively denotes the fact that someone tangibly read a tweet (not necessarily the linked content) among the sources they follow and is minimally interested in the topics evoked in that tweet. 
    \item We jointly consider interactional and informational attention. In this respect, focusing on retweets provides a uniform way to discuss social and semantic attention.
    In the case of semantic attention, we will nonetheless later show that results are consistent when considering all tweeting activities or just retweets: this further suggests that it remains sound to study both types of attention through retweets only.
\end{itemize}
}

\bigskip\noindent{\em Follower and retweet networks.}
To this end, we introduce two key networks, which share the same set of nodes (user accounts).
A user who is interested in the content published by another account may ``follow'' it, and thus subscribe to their posts, or ``tweets''. 
Over time, each user constitutes their own portfolio of such subscriptions: this defines the first network, the \emph{follower network}. 
Twitter exposes users to a portion of the tweets published by the accounts they follow, also denoted as ``followees''. 
Among this information feed, users may sometimes republish a post that they find particularly relevant, \hbox{i.e.} ``retweet'' it.
This creates the basis for the second network, the \emph{retweet network}, whose links denote the fact that a user retweeted a post that one of their followees previously published (be it an original post or already a retweet).

\begin{figure}[!ht]
\centering
\includegraphics[width=1.\linewidth]{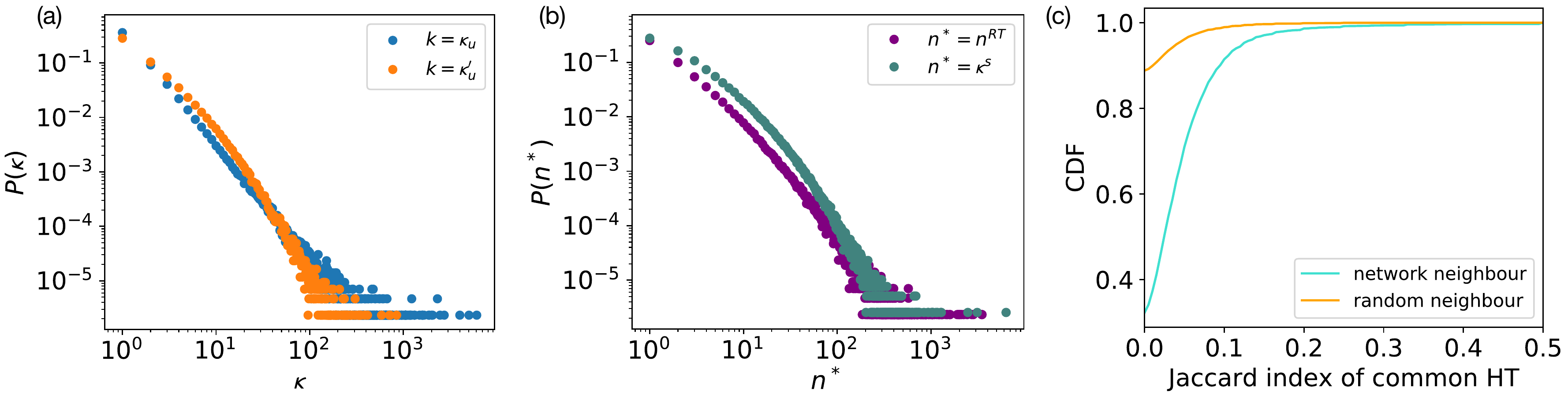}
\caption{Distributions of degrees, activity, and similarity measures. (a) Distributions of the $\kappa_u$ out- and $\kappa'_u$ in-degrees of nodes in the retweet network. (b) Distributions of the number of retweets ($n^{RT}$) and number of hashtags ($\kappa^s$) per user. (c) Cumulative distribution of the Jaccard similarity between hashtag sets of connected (blue) and randomly selected (orange) pairs of users of the retweet network.}
\label{fig:distributions}
\end{figure}

As such, a follower network denotes potential influence while a retweet network describes some form of actual influence: user retweets reveal successful exposure to content flows enabled by follower links.
Furthermore, a follower network may be defined in a static manner. While it evolves when users add (or remove) subscriptions at a given pace, at any time point $t$ a user still has a given and well-defined list of followees and followers. By contrast, a retweet network is fundamentally dynamic: it necessarily stems from the aggregation of observations of retweets over a certain time period and may only be defined by specifying a time range $[t,t']$.
More formally, we define: 
\begin{itemize}
\x the {\bf follower network $F_t$} at $t$ by adding a directed link $u\rightarrow v$ if $u$ follows $v$ at $t$, representing potential attention of $u$ to $v$ (as schematically shown in Fig.~\ref{fig:schema}a and b left panels). The out-degree $k_u$ of $u$ in that network directly denotes the number of followees of $u$, while the in-degree $k'_v$ denotes the number of followers of $v$.
\x the {\bf retweet network $R_{[t,t']}$} over $[t,t']$ by focusing on links $u\rightarrow v$ in $F_t$, then counting the number of times $u$ retweeted $v$'s tweets or retweeted a tweet after $v$ published that tweet, over the time period $[t,t']$ --- in what follows, this is precisely what we mean by ``retweet''. We add a weighted directed link $u\rightarrow v$ in $R_{[t,t']}$ with a weight $w_{uv}$ equal to that count (demonstrated in Fig.\ref{fig:schema}b right panels). The out-degree $\kappa_u$ denotes the number of users whom $u$ retweeted while the in-degree $\kappa'_v$ denotes the number of users who retweeted $v$. Distributions of these quantities are shown in Fig.~\ref{fig:distributions}a. The out-strength $s_u$ denotes the sum of the weights of the out-going links from $u$, \hbox{i.e.} number of retweets $u$ made of their followees, while the in-strength $s'_v$ denotes the total number of times $v$ has been retweeted by their followers.
\end{itemize}
Thus, links in $R_{[t,t']}$ form a subset of the links found in $F_t$. This implicitly relies on the assumption that $F_t$ provides a good approximation of $F_\tau$ at another time $\tau\in[t,t']$, which is indeed acceptable if $t'$ is sufficiently close to $t$ as will be the case in this paper: while the follower network seems to be highly dynamic in the long term, its evolution may be considered to be relatively limited over several weeks, where the proportion of replaced links remains in the vicinity of 10\% \citep{myers2014bursty}.

Figure~\ref{fig:schema} illustrates the construction process of both networks, where link directionality denotes some form of attention of the origin to the target.
From our data we constructed a directed follower network with $905,112$ customers as nodes and $69,156,298$ follower relationships as links.  The directed retweet network features $428,404$ nodes and $937,242$ links.

\subsection{Attentional degree}

Follower links thus denote a binary notion of potential attention: either one follows, or not.  By contrast, retweet links correspond to the magnitude of some form of actual attention. In other words, through retweets we observe how followers diversely allocate part of their attention to their followees.  Thus, for a given user $u$, the distribution of weights of their actual retweets of their followees $v_1,...,v_{\kappa_u}$ may be heterogeneous to some level: some users may allocate most of their attention to a certain followee, while others may balance their retweets across their whole portfolio of followees. To capture a notion of attention allocation, we use a measure of statistical dispersion, the Herfindahl-Hirschman index (HHI) that we apply on normalized weights of retweet (out-going) links from $u$:
\begin{equation}
H_u=\sum_i^{\kappa_u}{(\frac{w_{uv_i}}{s_u})}^2
\end{equation}
We call \emph{attentional degree} $a_u=1/H_u$, which varies between 1 (when attention is entirely focused on a single neighbor: $\exists v, w_{uv}=s_u$) and $\kappa_u$ (when attention is evenly split among all neighbors: $\forall i, w_{uv_i}=1/s_u$).
In economics, for instance, this value is computed on market shares of competing firms in a given market \citep{rhoades1993herfindahl}. It is interpreted as the number of ``equivalent firms'' in order to assess whether there is sufficient competition. If two firms, each having slightly less than 50\% of the market, although many other firms share the rest, this index is going to be close to two, indicating a duopoly in spite of an apparently high number of firms. In our context, $a_u$ represents by analogy the number of equivalent users whom $u$ devotes a \emph{meaningful} share of their attention to. We contend that this value realistically captures the number of neighbors with whom a node principally communicates in a directed weighted network. Note that, from this value, it would be possible to compute a reduced network by conserving edges with top weights in such a way that the number of neighbors of $u$ would coincide with $a_u$. It would however require us to additionally specify what to do with edges of equal weight close to the threshold induced by $a_u$. See also Appendix~\ref{sec:backbone} for a broader discussion on the meaning of this measure from the viewpoint of network reduction techniques.

\smallskip
We thus consider three layers of attention: first, the follower out-degree as the potential attention, second, the retweet out-degree as the actual attention, third, the attentional degree as the meaningful attention. We are aware that the observation of retweets and the way they are relayed within the follower network remains a partial proxy to describe attention dynamics that does not capture the whole picture of either influence or attention --- it misses information, which not only impacted users yet did not lead to observable retweets, but also which circulated through many other possible information channels.


\section{Social attention}\label{sec:social}

\subsection{Distribution of roles}

We provide an introductory outlook on the data we analyse and its various basic metrics on Fig.~\ref{fig:distributions}, in terms of hashtags, retweets or followers. In particular, the number of both hashtags or retweets per user, and the number of both followers or followees for each user all follow a typical and unsurprising heterogeneous distribution: many have little, few have a lot. We obtained equally unsurprising results when measuring the semantic similarity between pairs of users as the Jaccard index of their hashtag sets. As shown in Fig.~\ref{fig:distributions}c this similarity appears to be significantly higher for pairs of users connected in the retweet network, compared to random pairs.

\begin{figure*}[t]
\centering
\includegraphics[width=.9\linewidth]{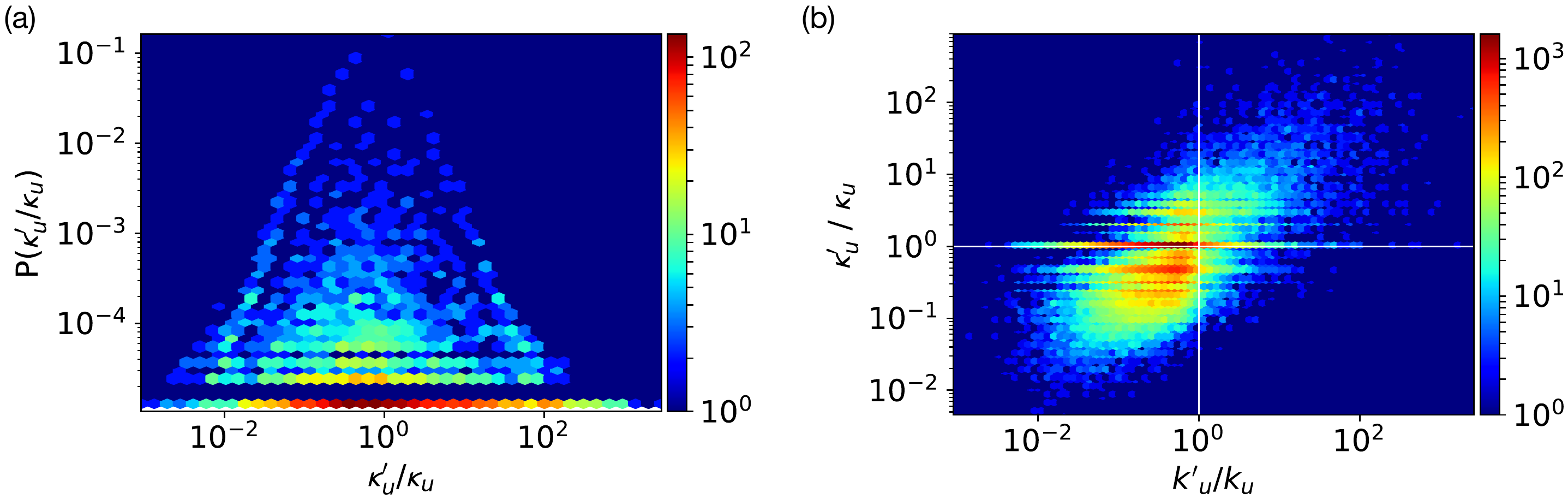}
\caption{Distributions of roles. (a) The $P(\kappa'_u/\kappa_u)$ distribution of retweet balance across all users shown as a density plot. (b) Configuration of the retweet balance (y-axis) in regard to the follower balance (x-axis), \hbox{i.e.} actual \hbox{vs.} potential attention flows measured as the correlation between the ratio of in- and out-degrees in the retweet and follower networks respectively. Heatmap colors code absolute counts.}
\label{fig:balanceratios}
\end{figure*}

In other words, on Twitter, like in many other online interaction contexts, we generally observe homophilic connection patterns while a significant part of the attention is concentrated on few users. Fig.~\ref{fig:balanceratios} focuses on the retweet network and further illustrates the {\em distribution of roles} in terms of potential input/output and actual input/output flows (measured as retweets). In particular, panel Fig.~\ref{fig:balanceratios}a shows the actual balance of flows as the $P(\kappa'_u/\kappa_u)$ distribution of the ratio between $\kappa'_u$ the number of times a user is retweeted and the $\kappa_u$ number of times a user retweets a neighbour in the retweet network. We call this the \emph{retweet balance} and show its distribution as a density plot due to discretization effect. We may similarly define the \emph{follower balance} as the ratio of followers \hbox{vs.} followees in the follower network. 

The right panel of Fig.~\ref{fig:balanceratios} then provides a broader picture by showing the density of users who exhibit a certain retweet balance (y-axis) with respect to their follower balance (x-axis), \hbox{i.e.} by comparing actual \hbox{vs.} potential attention flows in both directions.  This results in a double dichotomy that resembles what had been found earlier by \cite{gonzalez2013broadcasters}, even though they compared the follower with the mention networks. By contrast, we interpret this configuration from an attentional viewpoint, focusing on influence only and differentiating static from dynamic phenomena. In the top right quadrant, we find strong influencers who benefit from an excess of attention, both statically and dynamically; in the bottom left quadrant, what we may call normal users who pay static and dynamic attention to others rather than the other way around. The other two quadrants, top left and bottom right, are comparatively much less populated, they correspond respectively to users with a strong retweet balance yet a weak follower balance (actual flows are way above potential flows, in relative terms: so-called ``hidden influentials''), and to users with a weak retweet balance yet a strong follower balance (actual flows are way below potential flows, whom we may denote as ``fake influentials'').

These observations contribute to understand Twitter as a decentralized system of users who pay attention to others and forward information in a heterogeneous and somewhat homophilic manner. In this context, we now analyze the traces of these information dynamics to appraise whether cognitive limitations may nonetheless impose constraints on the functioning of this social information system.

\begin{figure*}
\centering
\includegraphics[width=.76\linewidth]{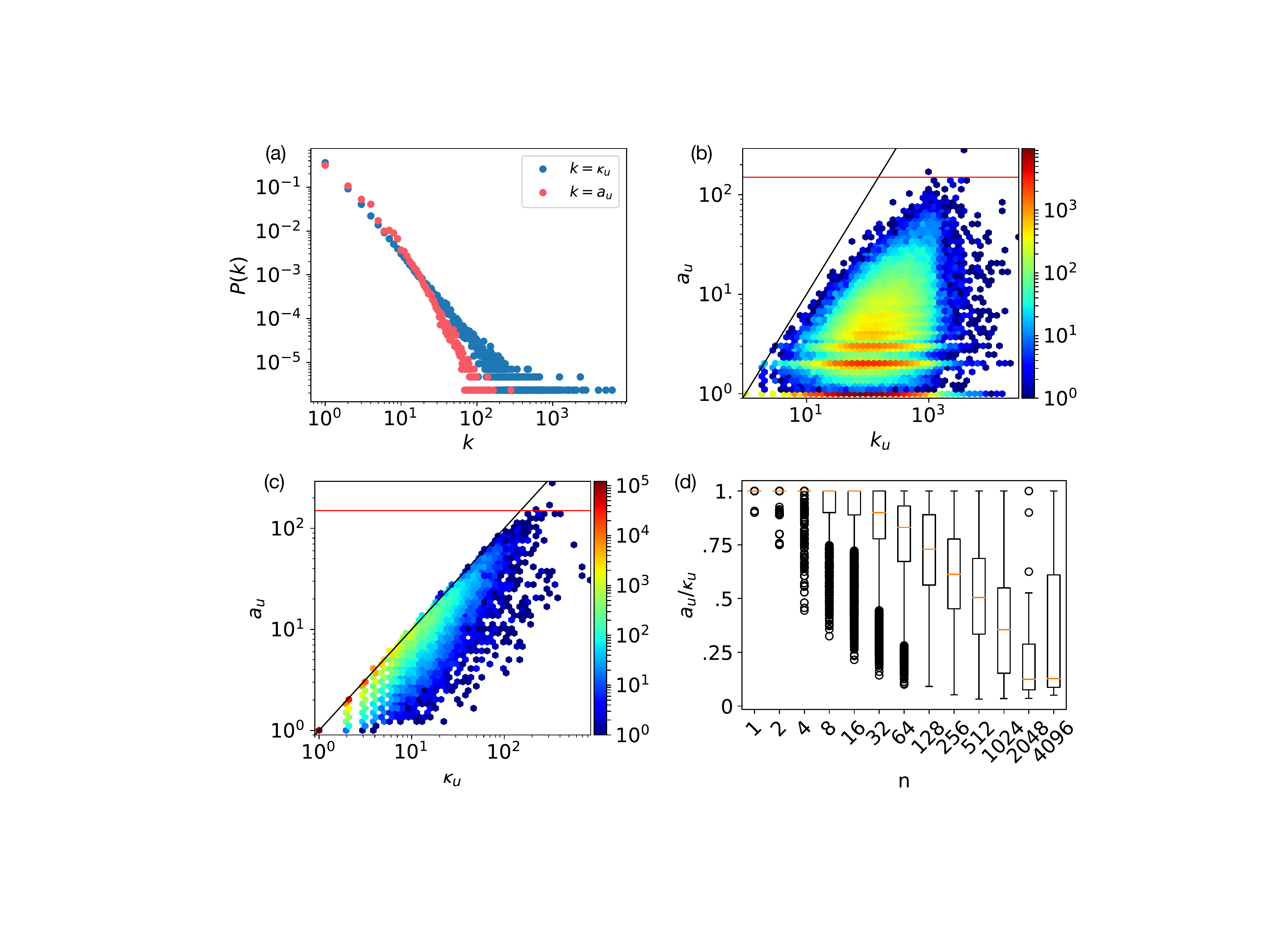}
\caption{(a) Distributions of the $a_u$ attentional degree and the $\kappa_u$ out-degree of the retweet network. (b) Attentional degree \hbox{vs.} out-degree in follower network. 
(c) Attentional degree \hbox{vs.} out-degree $\kappa_u$ in the retweet network. 
(d) Ratio between attentional degree and retweet out-degree $\kappa_u/k_u$ (y-axis) as a function of activity as the total $n$ number of tweets and retweets (x-axis). Heatmaps' colors code absolute counts.}
\label{fig:attall}
\end{figure*}

\subsection{Attention concentration}
To this end, we turn to the measure of meaningful attention, through the analysis of the attentional degree $a_u$ measured in our Twitter data. As shown in Fig.~\ref{fig:attall}a, this quantity appears to be heterogeneously distributed, but it takes values over a limited range as compared to the corresponding out-degree distribution of the retweet network. Next, to see how concentrated attention is, we plot the density of $a_u$ \hbox{vs.} $k_u$ in Fig.~\ref{fig:attall}b. This tells us to what extent users focus on a certain number of users among their followees. Irrespective of the number of followees, we observe that the attentional degree remains concentrated around a relatively narrow core in the order of about $\sim10$ users. For any user, it does not go above a hard threshold of a hundred users, even for users who follow hundreds or thousands of accounts. The observation that the reduced number of meaningful neighbours appears with such an upper limit is in a striking accordance with earlier suggestions. Dunbar's hypothesis suggests~\cite{roberts2009exploring,hill2003social} that, due to cognitive limitations, a person can maintain only about $150$ meaningful social relationships at the time (shown as a red solid line in Fig.~\ref{fig:attall}b and c). Further, other empirical studies, using different metrics than here, confirmed this suggestion in case of mobile phone communications~\cite{miri-limi} and in case of online social platforms, like Twitter~\cite{gonc-mode}, which would otherwise allow for more economic ways of interactions than traditional communication means.

We characterize further this core by comparing $a_u$ to $\kappa_u$. Deviations from the diagonal indicate to what extent the meaningful number of retweeted users ($a_u$) is smaller than their total number ($\kappa_u$). We plot this Fig.~\ref{fig:attall}c, which shows that the two values are somewhat related: density is generally higher close to diagonal. In other words, there is generally a good correlation between the raw number users who are given any attention to and the equivalent number of such users: once we pay attention to some users among our followees, attention appears to be relatively evenly distributed across them ($a_u\sim\kappa_u$). We propose to describe this configuration as two levels of attention, whereby user first pay most of their attention to a core of their followees while somewhat neglecting a periphery, and then equally pay attention to this core in weighted terms.

\medskip
\par\noindent{\em Two-level flows of attention.}
Furthermore, Fig.~\ref{fig:attall}c also exhibits some strong deviations, which indicate that some users further restrict their attention to an even smaller super-core of accounts. To understand better who these users are, we examine the ratio between $a_u$ and $\kappa_u$ and differentiate it with respect to increasing classes of activity, measured as $n=n^{TW}+n^{RT}$, i.e., the total number of tweets and retweets published by a user. Corresponding results in Fig~\ref{fig:attall}d seem to suggest that users with the highest level of activity correspond to those where the deviation \hbox{w.r.t.} the diagonal $y=x$ is strongest, \hbox{i.e.} where there is a higher loss/dissipation between attentional and retweet out-degrees. In other words, beyond a certain level of tweeting activity, only a core of users may receive significant attention, at least in relative terms. The above-described attentional thresholding is thus stronger for more active users, even though they naturally appear to pay attention to a higher raw number of users. This is consistent with similar observations on the allocation of attention on Facebook \citep{backstrom2011center} where the more active users still devote a large portion of their interactions to a small core of about 15 top users.

To summarize, we may describe these dynamics as ``{\em two-level flows of attention}'': first, users retweet \hbox{i.e.} focus on some followees only, among all potential followees (attention degree is capped with respect to the number of followees, but generally close to the out-degree in the retweet network); second, higher activity induces a stronger focus in relative terms or, put the other way around, higher activity does not really seem to enable a corresponding widening of attention toward a proportionally more varied set of users.  On the whole, this appears to point at the existence of a (converging) social lens which activity does not weaken much.


\section{Semantic attention}\label{sec:semantic}

\subsection{Semantic attentional degree}
\begin{figure}
\centering
\includegraphics[width=1.\linewidth]{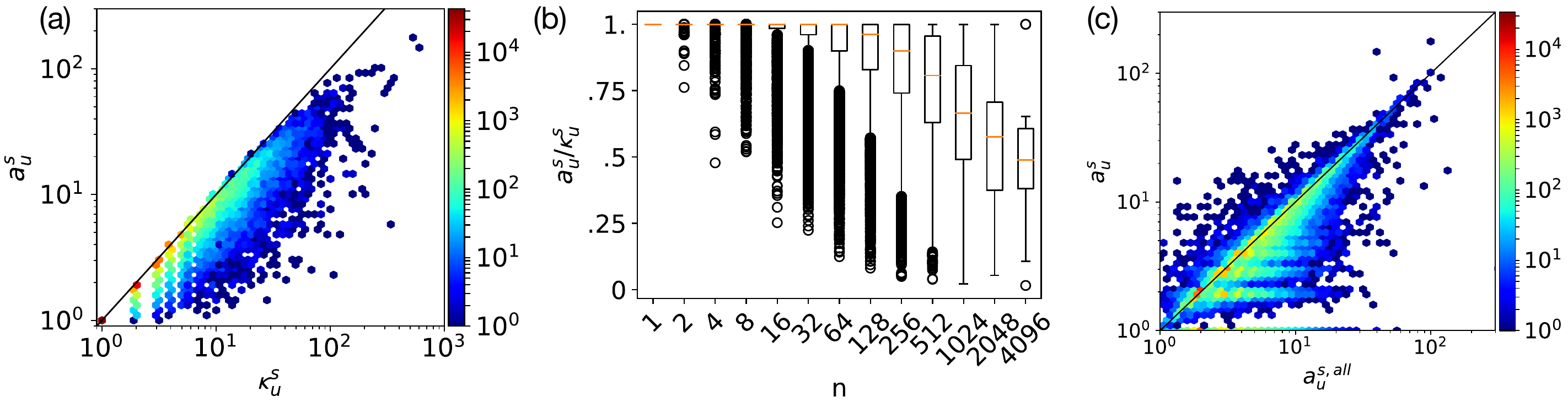}
\caption{\label{fig:semdegree} (a): Semantic attentional degree as a function of the hashtag set size. (b): Ratio between semantic attentional degree and number of hashtags $a^s_u/\kappa^s_u$ (y-axis) as a function of activity as the total $n$ number of tweets and retweets (x-axis). Heatmap colors code absolute counts. {(c) Correlation between semantic attentional degrees $a_u^s$ computed by considering retweeted hashtags, and $a_u^{s,all}$ computed by considering any tweeted or retweeted hashtag of a user $u$. This correlation appears with a Pearson coefficient $R=0.889$ (p-value<0.05).}}
\end{figure}

Users not only pay attention to some users but they may also focus on some issues: their posting activities may be devoted to a variety of topics, which may also be capped. Does informational attention exhibit similar features as interactional attention, is there a link between social attentional constraints and semantic ones? To appraise this, we measure semantic attention in terms of the diversity of hashtags used in a user's \tb{retweets}. 
For a given user $u$, we consider all their \tb{retweets} 
and compute the vector of occurrence of hashtags that they used strictly more than once (\hbox{i.e.} we ignore hapaxes for a given user level): $\omega_{uh}$ denotes the number of times a hashtag $h$ has been used by $u$ \tb{in their retweets}. We may compute the HHI on $\omega$ and thus the semantic attentional degree $a^s_u$ as the inverse, which provides an indication of the number of equivalent hashtags or topics addressed by $u$. \tb{Fig.~\ref{fig:semdegree}c exhibits a good correlation between $a^s_u$ and attentional degrees $a_u^{s,all}$ computed on all tweets, not only retweets and, more broadly, all figures of this section were also computed by considering tweets. They all yielded qualitatively similar results, indicating that tweet and retweet behaviors are generally consistent with one another as regards semantic attention.} 

We compare this with the raw number of hashtags ever addressed by $u$ \tb{in their retweets}, which we define as the semantic degree $\kappa^s_u$ by simple analogy with the social degree. 
Fig.~\ref{fig:semdegree}a features the distribution of $a^s$ \hbox{vs.} $\kappa^s$, which is similar to comparing the attentional social degree with the out-degree in the retweet network (even if the results are not as detailed as in the social case, where we can additionally distinguish potential attention from the follower network). Here too, we observe that: first, semantic attention is capped to several hundreds of hashtags in raw terms ($\kappa^s$), and to slightly above a hundred topics in equivalent terms ($a^s$); 
second, certain users do focus on hashtags in the sense that there is a more or less pronounced deviation between $\kappa^s$ and $a^s$. On Fig.~\ref{fig:semdegree}b we show that this dissipation is also strongly dependent on activity, perhaps in a less pronounced manner than in the social case: the semantic attention for the more active users (from an activity of about a hundred tweets) tends to exhibit a magnifying effect that corresponds at most to half the raw number of hashtags. In other words, active users both have broader interests but also display expertise patterns. On the semantic as well as the social sides, these effects exhibit a strong variance. It is likely that, on the whole, they indicate the existence of two distinct sub-populations of users: users, at all activity levels, who have a naturally narrow range of interests/interactions, and users, generally among the most active ones, who are broad yet remain quite focused.

\begin{figure}
\centering
\includegraphics[width=\linewidth]{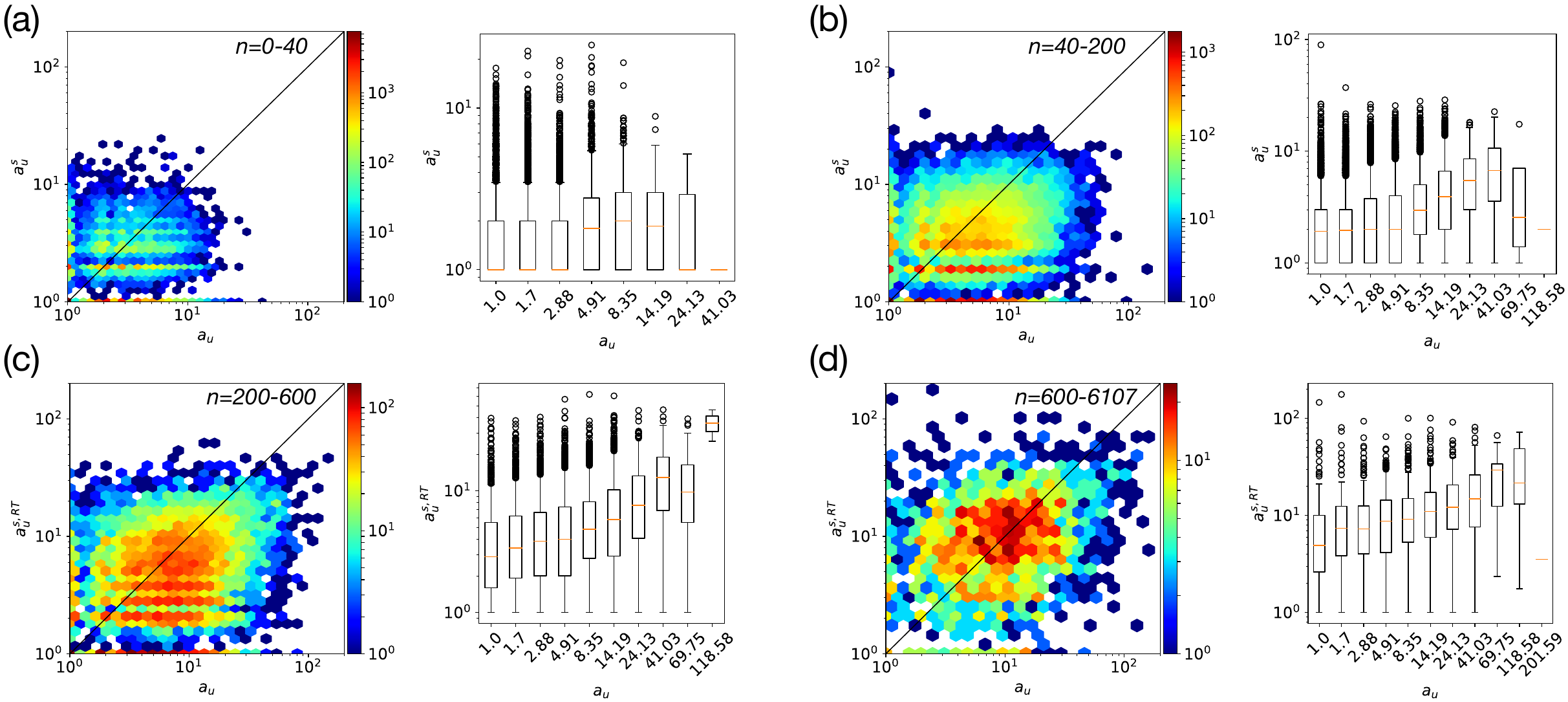}
\caption{\label{fig:semvssocrtw}Panels (a-d): semantic attention degree vs. social attention degree, split by the number of retweets. Left panels depicts correlation heatmaps between the semantic and social attention degrees of users belonging to a certain activity group. Right panels show the same information as box-plots. Activity $n$ of users are defined as the total number of their tweets and re-tweets.}
\end{figure}

\subsection{Socio-semantic correlations}

Cognitive constraints may thus apply on the interactional and informational sides. We now examine the combination of both and, in particular, aim to verify whether one has a link with the other. Two competing hypotheses may be proposed here. One is that of a joint reinforcement, or at least a positive correlation between both: users who pay attention to more actors also pay attention to more topics. The other hypothesis corresponds rather to a zero-sum game, where attention allocated to one dimension would likely constrain that allocated to the other.

In the former case, we may suggest that there exists an underlying activity variable (here, on Twitter), which would manifest itself in both dimensions: if users are able to cover more interactions, they also cover more topics, and vice-versa. This would be analogous to what has long been observed when comparing online and offline social capitals: while some have suggested that online sociability could deplete the potential for offline sociability, it has been shown that users who are socially more active online are also more active offline \citep{well-comp}
. In the latter case, \hbox{i.e.} a zero-sum game, we would observe a negative dependency between social and semantic attention, very much like what is behind Cobb-Douglas consumption graphs in economics where, for a constant level of possible consumption $C$ two possible goods $A$ and $B$ are consumed according to a function $C=A^\alpha B^{1-\alpha}$ such that consuming $A$ reduces $B$, generally in a non-linear fashion.

We plot the relationship between $a$ and $a^s$ on Fig.~\ref{fig:semvssocrtw}a-d, distinguishing various levels of $n$ posting activity (i.e., again, in terms of total publications). {For all levels, semantic attention appears to be correlated with social attention, with positive $R$ Pearson correlation coefficients and p-values summarised in Table~\ref{table:R}.} On heatmaps, highest densities are found around the diagonal, while boxplots again confirm a generally positive association between both types of attention.  Besides, as said before, there seems to be a non-linear relationship between posting activity and the average value of the social and semantic attentional degrees: centers of mass of degrees on all heatmaps do not move as quickly to the right as the center of mass for the respective retweet number ranges.

\setlength{\tabcolsep}{12pt}
\begin{table}[h!]
\centering
\begin{tabular}{ccccc}
 \toprule
 $n$ & $[1,40[$ & $[40,200[$ & $[200,600[$ & $[600,6107]$ \\
 \midrule
$R$ & 0.137 & 0.253 & 0.309 & 0.238 \\ 
$p$ & $<0.05$ & $<0.05$ & $<0.05$ & $<0.05$ \\ 
 \hline
\end{tabular}
\caption{\tb{Pearson correlation coefficient and p-value computed between the social and semantic attentional degrees of individuals in different $n$ retweeting activity groups.}}
\label{table:R}
\end{table}

On the whole, this seems to generally go in favor of the reinforcement hypothesis, moderated by posting activity in a non-linear fashion. Looking closely at all heatmaps, however, reveals that there is a bright horizontal (\hbox{resp.} vertical) band of high density of hexagons for small values of the vertical (\hbox{resp.} horizontal) axis, \hbox{i.e.} for a given small semantic attentional degree, for instance, there is a horizontal band of bright colored hexagons spanning several orders of magnitude of social attentional degree.  To summarize, there seems to be both a strong mass of points loosely around the diagonal, and a strong mass of points along vertical (\hbox{resp.} horizontal) lines for small values on the horizontal (\hbox{resp.} vertical) axis.  In other words, some users seem to focus exclusively either on the social side or the semantic users. It is unclear what the status of these users are (especially in terms of them being humans or bots) and this would warrant further research.


\section{Limitations}

\tb{Although we aimed to rely on some of the most general data filtering methods to obtain representative samples of Twitter activities, and on some of the least specific measures to capture attention, our study still has certain limitations. 
We wish to discuss some of them in this section to make it easier to draw more precise conclusions from our results.}  

\tb{First of all, we measured social attention via re-tweeting activity of a given neighbour. As discussed earlier, focusing on retweets may provide a partial view of attentional processes. Further, Twitter may offer other mechanisms to quantify more precisely these effects, for example by using \emph{likes}. However, this type of information was not available to us during the data collection period. Thus, we could not use them for a more precise quantification of attention. Attention signals may also be collected from external platforms: for instance, \citep{gabielkov2016social} uses audience data from \emph{bit.ly}, a link tracker website, to study the impact of shared links as a function of the number of followers of the users who posted them. While being a very sound protocol to study conversion rates, it also makes it difficult to match individual user characteristics across both datasets and thereby to discuss user-centric features, which are key from an attentional viewpoint.}

\tb{Second, Twitter does not only involve human actors but several fake accounts and robots, which may bias our observations.
If these non-human actors, or some of them, exhibit unrealistic posting and sharing activities, they may appear as outliers in our measurements --- assumedly, they would be unlikely to influence much the overall trends and core observations that we made on a larger population.}

\tb{Most of the data collected from a Twitter stream come as a sample restricted by pre-defined filters and collection rate limits. While filters are set up by the collector, rate limits induce some ambiguity in the data collection process. Collected data are commonly assumed to represent an unbiased sample of the Twitter stream coming as a fraction of tweets uniformly sampled from the set of tweets meeting the filtering conditions at a given time. While this assumption has been made in almost all Twitter studies, some work~\cite{morstatter2014biased} addressed and cautioned about the observational bias induced by the unknown sampler algorithm of Twitter. In the case of our dataset, we applied several language and location filters (as explained in Section~\ref{sec:dataset}) and obtained a relatively high rate of 15-25\% tweets via the PowerTrack API as compared to the Open API with only 1\% access. Despite this higher rate of data collection, which may considerably reduce the sampling bias we have in our data, we identify this ambiguity as a potential limitation, which is unfortunately present in the vast majority of other Twitter studies as well.}


\section{Discussion}\label{sec:discussion}

Twitter may be seen as a decentralized social information processing platform relying on users as input/output devices who are plugged onto their followees' information feeds, part of which they may or may not decide to dispatch to their followers. This decentralization is not devoid of hierarchy and heterogeneity.
From this viewpoint, at the collective level, it features a hierarchical yet roughly dichotomized distribution of roles: some users gather a lot of the potential and actual attention, many pay it, while potential and actual attention are generally correlated. Furthermore, at the individual level, we could hypothesize what we may call a ``two-level flow of attention'' whereby users first focus their actual attention on a core of their potential attention, then redistribute it in a relatively uniform way within that core. This observation was made possible by the use of a simple attentional focus measure, the attentional degree, which consists of a parameter-free approach to compute a number that may be easily be interpreted and compared with raw measures of numbers of neighbors in a network. On the whole, the limitations and focus effects that we find are consistent and, more importantly, extend the broad picture that has been depicted in other platforms in the literature. An interesting question that remains to be addressed would relate to the articulation between the collective and individual levels and, more precisely, to the description of the structural positions of the actors who gather the core of the attention of their neighbors, and the corresponding correlations. For instance, do the peers who are paid the most attention to \tb{at a user-centric level also} occupy certain positions in the network, {do they act as opinion leaders as suggested by Lazarsfeld's two step flow of communication hypothesis~\cite{berelson1968people,katz2017personal}}, and are their topological properties correlated among each other? This would shed light on the possible existence of a higher attention likelihood for some kinds of peers who are more likely to pass information and who may be found in specific parts of the network (e.g., in terms of distant communities, cohesive clusters, and so-called hubs and bridges)~\cite{weng2018attention}.

Finally, we completed the understanding of the constraints that apply to individual information processing by adopting a joint interactional and informational perspective. 
In particular, our last figure sheds light on a major question regarding the supply of attention: do semantic and social activities share the same limited supply and thus have a negative impact on one another (convex relationship between both) or do they actually reinforce one another (positive correlation along the diagonal) while being a sublinear function of the activity level? In this respect, we observed a relationship between social and semantic attentional processes whereby they are also generally correlated: we could demonstrate that for most users, both types of attentional resources are related in a positive manner. This hints at the existence of a heterogeneous distribution of attentional resources among users that expresses itself jointly on the semantic and social side, even though there are sometimes marked discrepancies between both types of attention (in terms of divergence \hbox{w.r.t.} the diagonal ``social attention=semantic attention'') and that there even exists a special minority of users who are exclusively focused on one side only (\hbox{i.e.} either semantic or social). Here again, a further research direction could consist in appraising the structural positions occupied by these very users --- especially by qualifying which users exhibit more social than semantic attention and whether they possess specific topological properties and populate specific parts of the whole system. This, together with the above-mentioned phenomena regarding the uneven distribution of attention, would be likely beneficial to influence studies and contribute to develop finer models that take into account the differential balance of attention among users and towards some selected peers.





\vspace{6pt} 



\authorcontributions{All authors participated in the design of the presented research and in the drafting the publication. A.B. performed the exploratory data analysis. M.K performed the advanced data analysis. C.R. and M.K. wrote the manuscript.}

\funding{This paper has been partially realized in the framework of the ``Algodiv'' grant (ANR-15-CE38-0001) funded by the ANR (French National Agency of Research). It was also supported by the SoSweet ANR project (ANR-15-CE38-0011), the ACADEMICS grant of the IDEXLYON, project of the Université de Lyon, PIA operated by ANR-16-IDEX-0005, and the ``Socsemics'' Consolidator grant funded by the European Research Council (ERC) under the European Union's Horizon 2020 research and innovation programme (grant agreement No. 772743).}

\acknowledgments{We are grateful for the data collection and preparation to J.-Ph. Magu\'e.}

\conflictsofinterest{The authors declare no conflict of interest. The funders had no role in the design of the study; in the collection, analyses, or interpretation of data; in the writing of the manuscript, or in the decision to publish the results.} 


%

\appendix 

\section{Backbone networks and attentional degrees}\label{sec:backbone}

Network backbones~\cite{Serrano6483} provide a filtering method (called the disparity filter) to extract the relevant connection backbone in weighted networks, keeping edges which represent statistically significant deviations for the local assignment of link weights as compared to a null model. In our case we refer to them as a reference system to show that our network reduction method does not evidently capture the same information and thus provides a novel way to identify important ties in a weighted network.

Taking the normalised weight $p_{ij}=w_{ij}/\sum_j w_{ij}$ of links connected to node $i$ we compare their values to a null model where $p_{ij}$ is sampled random uniform distribution of the total strength $\sum_j w_{ij}$ of the given node $i$. This sampling process corresponds to placing $k-1$ points uniformly over the period of $[0,1]$ which provides us $k$ sub-interval, where $k$ being the degree of node $i$. The probability distribution of the value $x$ of each of these intervals~\cite{Serrano6483} is
\begin{equation}
\rho(x)dx=(k-1)(1-x)^{k-1}dx.
\end{equation}
The disparity filter defines a probability $\alpha_{ij}$ for each link to decide whether it is compatible with the null hypothesis. It is equal to the probability that one of the $k$ intervals defined in the null model has a length larger than $p_{ij}$, that is
\begin{equation}
\alpha_{ij}=\int_{p_{ij}}^{1}\rho(x)dx=(1-p_{ij})^{k-1}.
\end{equation}
In case the link appears with $\alpha_{ij}<\alpha$, the null hypothesis rejected at the significance level of $\alpha$ and the link is kept in the filtered network. The significance parameter $\alpha$ is the only parameter of the filter and controls the number of links retained in the retained network. Note, that in case $k=1$ for a given node, its single link is automatically preserved.

Here we apply the disparity filter on the directed weighted retweet network to obtain the backbone structure of incoming links, which is corresponding to the reduced attention network of the same structure. This means~\cite{Serrano6483} that while computing $p_{ij}$ we only consider weights of incoming links of node $i$ and in turn $\alpha_{ij}\neq \alpha_{ji}$.

\begin{figure*}[ht]
\centering
\includegraphics[width=1.\linewidth]{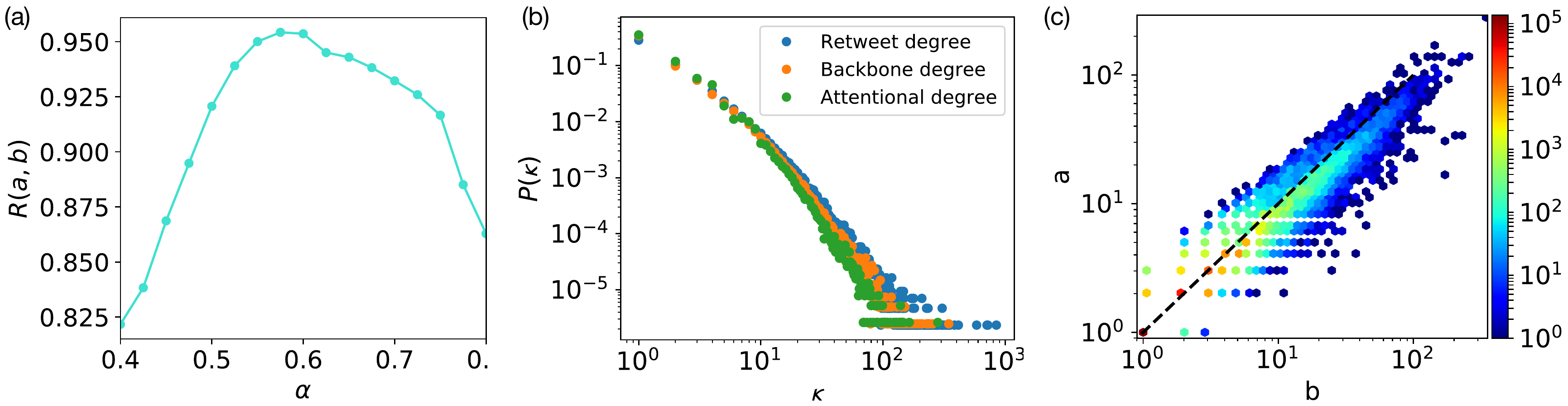}
\caption{Distributions and correlations of degrees measured in the backbone networks or as  attentional degrees. (a) $R(a,b)$ Pearson correlation values between the $a$ attentional and $b$ backbone degrees as the function of the $\alpha$ disparity filter parameter. (b) Distributions of retweet, backbone and attentional degrees of users. (c) Correlations between the $a$ attentional and $b$ backbone degrees computes with the optimal disparity filter value $\alpha=0.575$.}
\label{fig:A1}
\end{figure*}

In order to compare the backbone network and attentional degrees obtained from the retweet network, first we look for the case when they are the most similar in terms of their degrees. To find the best correspondence, we take the attentional degrees and compute the correlation with degrees of several backbone networks obtained by varying the $\alpha$ disparity filter parameter. The highest correlations has been found for $\alpha=0.575$ as shown Fig.~\ref{fig:A1}a, where we plot the $R(k_{in}^{att},k_{in}^{bb})$ Pearson correlation coefficient between node degrees in the two structures as the function of $\alpha$. Note that all observed correlations are significant with $p<0.05$. Comparing the degree distribution of the retweet network to the similar filtered graphs (see Fig.~\ref{fig:A1}b) we find that although they are very similar, the attention filter provides a network with the most reduced degree heterogeneities. To directly observe degree correlations, we show the degrees of nodes in the two selected filtered structures as a heat-map of a scatter plot in Fig.~\ref{fig:A1}c. There, despite the strong correlation between the various degree values ($R=0.955$, $p<0.05$), we find strong fluctuations as well, indicating that the two filtering process do not identify the same set of links to be important, which underlies the relevance of our method. Note that while our method is parameterless, disparity filter has a parameter, what we tuned to obtain the most similar structure. Without this tuning our method may provide even more different filtered set of links, which may hold different roles in the structure. Inversely, attentional degrees can be used to identify the optimal $\alpha$ disparity filter parameter without looking at more complicated network characteristics, as it was suggested originally~\cite{Serrano6483}.


\reftitle{References}

\begin{thebibliography}{-------}
\providecommand{\natexlab}[1]{#1}

\bibitem[Hill and Dunbar(2003)]{hill2003social}
Hill, R.A.; Dunbar, R.I.
\newblock Social network size in humans.
\newblock {\em Human nature} {\bf 2003}, {\em 14},~53--72.

\bibitem[Roberts \em{et~al.}(2009)Roberts, Dunbar, Pollet, and
  Kuppens]{roberts2009exploring}
Roberts, S.G.; Dunbar, R.I.; Pollet, T.V.; Kuppens, T.
\newblock Exploring variation in active network size: Constraints and ego
  characteristics.
\newblock {\em Social Networks} {\bf 2009}, {\em 31},~138--146.

\bibitem[Burke \em{et~al.}(2011)Burke, Kraut, and Marlow]{burke2011social}
Burke, M.; Kraut, R.; Marlow, C.
\newblock Social capital on Facebook: Differentiating uses and users.
\newblock  Proceedings of the SIGCHI conference on human factors in computing
  systems. ACM,  2011, pp. 571--580.

\bibitem[Miritello \em{et~al.}(2013)Miritello, Lara, Cebrian, and
  Moro]{miri-limi}
Miritello, G.; Lara, R.; Cebrian, M.; Moro, E.
\newblock Limited communication capacity unveils strategies for human
  interaction.
\newblock {\em Scientific Reports} {\bf 2013}, {\em 3},~1--7.

\bibitem[Weng \em{et~al.}(2012)Weng, Flammini, Vespignani, and
  Menczer]{weng-comp}
Weng, L.; Flammini, A.; Vespignani, A.; Menczer, F.
\newblock Competition among memes in a world with limited attention.
\newblock {\em Scientific Reports} {\bf 2012}, {\em 2},~335.

\bibitem[Wellman(2001)]{well-comp}
Wellman, B.
\newblock Computer Networks as Social Networks.
\newblock {\em Science} {\bf 2001}, {\em 293},~2031--2034.

\bibitem[Wellman \em{et~al.}(2001)Wellman, Haase, Witte, and
  Hampton]{wellman-internet-increase-2001}
Wellman, B.; Haase, A.Q.; Witte, J.; Hampton, K.
\newblock Does the Internet Increase, Decrease, or Supplement Social Capital?:
  Social Networks, Participation, and Community Commitment.
\newblock {\em American Behavioral Scientist} {\bf 2001}, {\em 45},~436--455.

\bibitem[Orbach \em{et~al.}(2015)Orbach, Demko, Doyle, Waber, and
  Pentland]{orbach-2015-sensing}
Orbach, M.; Demko, M.; Doyle, J.; Waber, B.N.; Pentland, A.S.
\newblock Sensing Informal Networks in Organizations.
\newblock {\em American Behavioral Scientist} {\bf 2015}, {\em 59},~508--524.

\bibitem[Nguyen and Lethiais(2016)]{nguyen-2016-the-impact}
Nguyen, G.D.; Lethiais, V.
\newblock The Impact of Social Networks on Sociability : The Case of Facebook.
\newblock {\em R\'eseaux} {\bf 2016}, {\em 195},~165--195.

\bibitem[Newman(2001)]{newm:str1}
Newman, M.E.J.
\newblock Scientific collaboration networks. {I}. {N}etwork construction and
  fundamental results.
\newblock {\em Physical {R}eview {E}} {\bf 2001}, {\em 64},~016131.

\bibitem[Ebel \em{et~al.}(2002)Ebel, Mielsch, and Bornholdt]{ebel2002scale}
Ebel, H.; Mielsch, L.I.; Bornholdt, S.
\newblock Scale-free topology of e-mail networks.
\newblock {\em Physical review E} {\bf 2002}, {\em 66},~035103.

\bibitem[Kumar \em{et~al.}(2010)Kumar, Novak, and Tomkins]{kumar2010structure}
Kumar, R.; Novak, J.; Tomkins, A.
\newblock Structure and evolution of online social networks. In {\em Link
  mining: models, algorithms, and applications}; Springer,  2010; pp. 337--357.

\bibitem[Mislove \em{et~al.}(2007)Mislove, Marcon, Gummadi, Druschel, and
  Bhattacharjee]{mislove2007measurement}
Mislove, A.; Marcon, M.; Gummadi, K.P.; Druschel, P.; Bhattacharjee, B.
\newblock Measurement and analysis of online social networks.
\newblock  Proceedings of the 7th ACM SIGCOMM conference on Internet
  measurement. ACM,  2007, pp. 29--42.

\bibitem[Gon{\c c}alves \em{et~al.}(2011)Gon{\c c}alves, Perra, and
  Vespignani]{gonc-mode}
Gon{\c c}alves, B.; Perra, N.; Vespignani, A.
\newblock Modeling Users' Activity on Twitter Networks: Validation of Dunbar's
  Number.
\newblock {\em {PLoS} {ONE}} {\bf 2011}, {\em 6},~e22656.

\bibitem[Lehmann \em{et~al.}(2012)Lehmann, Gon{\c c}alves, Ramasco, and
  Cattuto]{lehm-dyna}
Lehmann, J.; Gon{\c c}alves, B.; Ramasco, J.; Cattuto, C.
\newblock Dynamical Classes of Collective Attention in Twitter.
\newblock  Proceeding 21st WWW'12 Intl Conf on World Wide Web,  2012, pp.
  251--260.

\bibitem[Sanl{\i} and Lambiotte(2015)]{sanli2015local}
Sanl{\i}, C.; Lambiotte, R.
\newblock Local variation of hashtag spike trains and popularity in twitter.
\newblock {\em PloS one} {\bf 2015}, {\em 10},~e0131704.

\bibitem[Backstrom \em{et~al.}(2011)Backstrom, Bakshy, Kleinberg, Lento, and
  Rosenn]{backstrom2011center}
Backstrom, L.; Bakshy, E.; Kleinberg, J.M.; Lento, T.M.; Rosenn, I.
\newblock Center of attention: How facebook users allocate attention across
  friends.
\newblock  Fifth International AAAI Conference on Weblogs and Social Media,
  2011.

\bibitem[Saram{\"a}ki \em{et~al.}(2014)Saram{\"a}ki, Leicht, L{\'o}pez,
  Roberts, Reed-Tsochas, and Dunbar]{saramaki2014persistence}
Saram{\"a}ki, J.; Leicht, E.A.; L{\'o}pez, E.; Roberts, S.G.; Reed-Tsochas, F.;
  Dunbar, R.I.
\newblock Persistence of social signatures in human communication.
\newblock {\em Proceedings of the National Academy of Sciences} {\bf 2014},
  {\em 111},~942--947.

\bibitem[Lin \em{et~al.}(2014)Lin, Keegan, Margolin, and Lazer]{lin2014rising}
Lin, Y.R.; Keegan, B.; Margolin, D.; Lazer, D.
\newblock Rising tides or rising stars?: Dynamics of shared attention on
  Twitter during media events.
\newblock {\em PloS one} {\bf 2014}, {\em 9},~e94093.

\bibitem[Golder and Huberman(2006)]{golder-2006-usage}
Golder, S.A.; Huberman, B.A.
\newblock Usage patterns of collaborative tagging systems.
\newblock {\em Journal of Information Science} {\bf 2006}, {\em 32},~198--208.

\bibitem[Wu and Huberman(2007)]{wu-2007-nov}
Wu, F.; Huberman, B.A.
\newblock Novelty and collective attention.
\newblock {\em {PNAS}} {\bf 2007}, {\em 104},~17599--17601.

\bibitem[Weng \em{et~al.}(2013)Weng, Menczer, and Ahn]{weng-vira}
Weng, L.; Menczer, F.; Ahn, Y.Y.
\newblock Virality Prediction and Community Structure in Social Networks.
\newblock {\em Scientific Reports} {\bf 2013}, {\em 3},~2522.

\bibitem[Romero \em{et~al.}(2011)Romero, Meeder, and Kleinberg]{rome-diff}
Romero, D.M.; Meeder, B.; Kleinberg, J.
\newblock Differences in the Mechanics of Information Diffusion Across Topics:
  Idioms, Political Hashtags, and Complex Contagion on Twitter.
\newblock  Proc. {ACM} WWW'11 Mar 28-Apr 1, 2011,  2011.

\bibitem[Hodas and Lerman(2012)]{hodas2012visibility}
Hodas, N.O.; Lerman, K.
\newblock How visibility and divided attention constrain social contagion.
\newblock  2012 International conference on Privacy, Security, Risk and Trust
  (PASSAT), and 2012 International Conference on Social Computing (SocialCom).
  IEEE,  2012, pp. 249--257.

\bibitem[Rodriguez \em{et~al.}(2014)Rodriguez, Gummadi, and
  Schoelkopf]{rodriguez2014quantifying}
Rodriguez, M.G.; Gummadi, K.; Schoelkopf, B.
\newblock Quantifying information overload in social media and its impact on
  social contagions.
\newblock  Eighth International AAAI Conference on Weblogs and Social Media,
  2014.

\bibitem[Unicomb \em{et~al.}(2018)Unicomb, I{\~n}iguez, and
  Karsai]{unicomb2018threshold}
Unicomb, S.; I{\~n}iguez, G.; Karsai, M.
\newblock Threshold driven contagion on weighted networks.
\newblock {\em Scientific reports} {\bf 2018}, {\em 8},~3094.

\bibitem[Suh \em{et~al.}(2010)Suh, Hong, Pirolli, and Chi]{suh2010want}
Suh, B.; Hong, L.; Pirolli, P.; Chi, E.H.
\newblock Want to be retweeted? Large-Scale analytics on factors impacting
  retweet in twitter network.
\newblock  2010 IEEE Second International Conference on Social Computing. IEEE,
   2010, pp. 177--184.

\bibitem[Yang \em{et~al.}(2010)Yang, Guo, Cai, Tang, Li, Zhang, and
  Su]{yang2010understanding}
Yang, Z.; Guo, J.; Cai, K.; Tang, J.; Li, J.; Zhang, L.; Su, Z.
\newblock Understanding retweeting behaviors in social networks.
\newblock  Proceedings of the 19th ACM International Conference on Information
  and knowledge management. ACM,  2010, pp. 1633--1636.

\bibitem[Zaman \em{et~al.}(2010)Zaman, Herbrich, Van~Gael, and
  Stern]{zaman2010predicting}
Zaman, T.R.; Herbrich, R.; Van~Gael, J.; Stern, D.
\newblock Predicting Information Spreading in Twitter.
\newblock  Computational Social Science and the Wisdom of Crowds Workshop
  (colocated with NIPS 2010),  2010.

\bibitem[boyd \em{et~al.}(2010)boyd, Golder, and Lotan]{boyd2010tweet}
boyd, D.; Golder, S.; Lotan, G.
\newblock Tweet, tweet, retweet: Conversational aspects of retweeting on
  twitter.
\newblock  2010 43rd Hawaii International Conference on System Sciences. IEEE,
  2010, pp. 1--10.

\bibitem[Myers and Leskovec(2014)]{myers2014bursty}
Myers, S.A.; Leskovec, J.
\newblock The bursty dynamics of the twitter information network.
\newblock  Proceedings of the 23rd international conference on World wide web.
  ACM,  2014, pp. 913--924.

\bibitem[Rhoades(1993)]{rhoades1993herfindahl}
Rhoades, S.A.
\newblock The herfindahl-hirschman index.
\newblock {\em Fed. Res. Bull.} {\bf 1993}, {\em 79},~188.

\bibitem[Gonz{\'a}lez-Bail{\'o}n \em{et~al.}(2013)Gonz{\'a}lez-Bail{\'o}n,
  Borge-Holthoefer, and Moreno]{gonzalez2013broadcasters}
Gonz{\'a}lez-Bail{\'o}n, S.; Borge-Holthoefer, J.; Moreno, Y.
\newblock Broadcasters and hidden influentials in online protest diffusion.
\newblock {\em American Behavioral Scientist} {\bf 2013}, {\em 57},~943--965.

\bibitem[Gabielkov \em{et~al.}(2016)Gabielkov, Ramachandran, Chaintreau, and
  Legout]{gabielkov2016social}
Gabielkov, M.; Ramachandran, A.; Chaintreau, A.; Legout, A.
\newblock Social clicks: What and who gets read on Twitter?
\newblock {\em ACM SIGMETRICS Performance Evaluation Review} {\bf 2016}, {\em
  44},~179--192.

\bibitem[Morstatter \em{et~al.}(2014)Morstatter, Pfeffer, and
  Liu]{morstatter2014biased}
Morstatter, F.; Pfeffer, J.; Liu, H.
\newblock When is it biased?: assessing the representativeness of twitter's
  streaming API.
\newblock  Proceedings of the 23rd international conference on world wide web.
  ACM,  2014, pp. 555--556.

\bibitem[Berelson \em{et~al.}(1968)Berelson, Gaudet, and
  Lazarsfeld]{berelson1968people}
Berelson, B.; Gaudet, H.; Lazarsfeld, P.F.
\newblock {\em The people's choice: How the voter makes up his mind in a
  presidential campaign}; Columbia University Press,  1968.

\bibitem[Katz \em{et~al.}(2017)Katz, Lazarsfeld, and Roper]{katz2017personal}
Katz, E.; Lazarsfeld, P.F.; Roper, E.
\newblock {\em Personal influence: The part played by people in the flow of
  mass communications}; Routledge,  2017.

\bibitem[Weng \em{et~al.}(2018)Weng, Karsai, Perra, Menczer, and
  Flammini]{weng2018attention}
Weng, L.; Karsai, M.; Perra, N.; Menczer, F.; Flammini, A.
\newblock Attention on weak ties in social and communication networks. In {\em
  Complex Spreading Phenomena in Social Systems}; Springer,  2018; pp.
  213--228.

\bibitem[Serrano \em{et~al.}(2009)Serrano, Bogu{\~n}{\'a}, and
  Vespignani]{Serrano6483}
Serrano, M.{\'A}.; Bogu{\~n}{\'a}, M.; Vespignani, A.
\newblock Extracting the multiscale backbone of complex weighted networks.
\newblock {\em Proceedings of the National Academy of Sciences} {\bf 2009},
  {\em 106},~6483--6488,
  \href{http://xxx.lanl.gov/abs/https://www.pnas.org/content/106/16/6483.full.pdf}{{\normalfont
  [https://www.pnas.org/content/106/16/6483.full.pdf]}}.
\newblock
  doi:{\changeurlcolor{black}\href{https://doi.org/10.1073/pnas.0808904106}{\detokenize{10.1073/pnas.0808904106}}}.

\end{thebibliography}
\end{document}